\documentclass{webofc}
\usepackage{color, colortbl}
\usepackage[dvipsnames]{xcolor}

\usepackage[varg]{txfonts}   
\newcommand{\xmm}{{\it XMM-Newton}}
\begin{document}
\title{On the dynamical and morphological state of the CHEX-MATE clusters}
%
%

\author{\firstname{Maria Giulia} \lastname{Campitiello}\inst{1,2}\fnsep\thanks{\email{giulia.campitiello@inaf.it}} \and 
	\firstname{Stefano} \lastname{Ettori}\inst{1,3} 
	\and
	\firstname{Lorenzo} \lastname{Lovisari}\inst{1,4} and the CHEX-MATE collaboration
}

\institute{INAF, Osservatorio di Astrofisica e Scienza dello Spazio, via Piero Gobetti 93/3, 40129 Bologna, Italy
         \and
Dipartimento di Fisica e Astronomia, Università di Bologna, Via Gobetti 92/3, 40121, Bologna, Italy
         \and
INFN, Sezione di Bologna, viale Berti Pichat 6/2, 40127 Bologna, Italy
         \and 
Center for Astrophysics $|$ Harvard $\&$ Smithsonian, 60 Garden Street, Cambridge, MA 02138, USA
          }

\abstract{%
The CHEX-MATE sample was built to provide an overview of the statistical properties of the underlying cluster population and to set the stage for future X-ray missions. In this work, we perform a morphological analysis of the 118 clusters included in the sample with the aim to provide a classification of their dynamical state which will be useful for future studies of the collaboration.}
\maketitle
\section{Introduction} \label{intro}
Clusters of galaxies are the largest virialised systems of the Universe and include between 100 - 1000 galaxies. However, the galactic and star component represents only the 15\% of their barionic mass. The bulk of the baryons resides in the Intra Cluster Medium (ICM), a hot ionised plasma that emits in the X-ray band. 
By identifying specific features in the X-ray morphology of clusters is possible to derive their dynamical state, which is an important property to consider in both astrophysical and cosmological studies. Clusters show a large variety of dynamical states, but we can essentially identify two extreme classes of objects. On the one hand there are the relaxed systems, whose X-ray emission is characterised by a spherical symmetry and are assumed to be in hydrostatic equilibrium allowing a proper mass estimate \citep[e.g.,][]{Ettori2013,Pratt2019}, which is an important cosmological quantity. 
On the other hand, there are the disturbed clusters which show clear signs of mergers and which are useful for astrophysical studies related, for example, to turbulence or particle re-acceleration mechanisms \citep[e.g.,][]{Cassano2010, Mann2012}. Therefore, a classification of the dynamical state is crucial when dealing with large samples of clusters, since it allows to recognise the most proper objects to use in specific analysis. One of the most recent samples of clusters observed in the X-ray band is the CHEX-MATE sample \citep[the Cluster HEritage project with XMM-Newton Mass Assembly and Thermodynamics at the Endpoint of structure formation; see][]{chexmate21}, which is a 3 Ms Multi-Year XMM-Newton Heritage Programme to obtain X-ray observations of a minimally-biased, signal-to-noise-limited sample of clusters detected by Planck through the Sunyaev–Zeldovich effect (SZE).
The goals of the project are to provide an accurate vision of the statistical properties of the underlying population, measure how the gas properties are shaped by collapse into the dark matter halo, uncover the provenance of non-gravitational heating, and resolve the major uncertainties in mass determination that limit the use of clusters for cosmological studies. In this work, we present the morphological analysis of the clusters in the CHEX-MATE sample with the aim to provide a classification of their dynamical state, which will be useful for future studies of the collaboration. Furthermore, we will retrace the steps taken over the last years to test and check the procedure elaborated until now in this field.

\section{Dataset and methods}\label{sec-1}
The CHEX-MATE\footnote{http://xmm-heritage.oas.inaf.it/} sample consists of 118 clusters detected by Planck at high signal-to-noise (S/N> 6.5) through their SZE signal, spanning a wide range of masses and redshifts. The sample is divided in two groups: the $Tier$ $1$, consisting of 61 objects located at low redshift (0.05 < z < 0.2) in the Northern sky (DEC > 0) with a total mass of 2 $\times$ 10$^{14}$ M$_{\odot}<$ M$_{500}<$ 9 $\times$ 10$^{14}$ M$_{\odot}$, providing an unbiased view of the population at the most recent time; the \textit{Tier 2}, including the most massive systems to have formed thus far in the history of the Universe (z < 0.6 with M$_{500}$ > 7.25 $\times$ 10$^{14}$ M$_{\odot}$). Four clusters are in common between these two sub-samples. The XMM-Newton X-ray images (FWHM of the PSF $\sim$ 6 arcsec) used in this analysis are filtered in the in the [0.7-1.2 keV] band, background-subtracted, and exposure corrected. All point-sources were masked to avoid any contamination of our results. Finally, the signal to noise ratio that characterised the images is S/N$\sim$150, for most of the objects, providing a uniform characterisation of the entire sample.
\subsection{Visual classification}
As first step of our study, we considered the oldest method used to classify the clusters X-ray morphology, i.e. the visual classification. In particular, 7 astronomers inspected the cluster images by eye, and assigned a grade from 0 to 3 to each object, where 0 are the most relaxed systems and 3 the most disturbed ones. The results of the 7 classifications were then averaged and clusters with rounded values equal to 0 were classified as relaxed (R, 18 clusters), clusters with rounded values equal to 3 as disturbed (D, 32 clusters), and all the other clusters as mixed (M, 68 clusters). The visual classification is a very subjective and time-consuming procedure and it is not suitable for large samples of objects or for future surveys. However, in our case, the involvement of 7 people allows to reduce biases related to the subjectivity of the method, and consequently we will use the results obtained as reference for the rest of the analysis.
\subsection{Morphological parameters} \label{Sec-morpho}
The limits of the visual classification described above led to the definition of more robust indicators able to objectively quantify even small deviations  from  a  perfectly regular  and  spherically-symmetric emission. Below we report the parameters considered in this work:
\begin{itemize}
    \item the concentration, $c$, defined as the ratio of the surface brightness (SB) inside two concentric apertures \citep{Santos2008}: $c = SB_{(r < 0.15\text{ R$_{500}$})} / SB_{(\text{r}<\text{R$_{500}$})}$;

    \item the centroid shift, $w$, \citep{Mohr1995} defined as the standard deviation of the projected separation between the X-ray peak and the centroid of the emission computed within N (=10 in our case) apertures of increasing radius: $w=\frac{1}{\text{R}_{500}} \sqrt{ \frac{1}{\text{N}-1}\sum_i (\Delta_i-\Bar{\Delta})^2 }$
where $\Delta_i$ is the distance between the X-ray peak and the centroid of the i-th aperture, and R$_{500}$ is the radius of the largest aperture. 
    \item the power ratios \citep{Buote1995}, which are a multipole decomposition of the X-ray surface brightness inside a certain aperture. The $m$-order ($m>$0) power ratio is defined as P$_m$/P$_{0}$, with: $P_0=[a_0 \text{ln}(\text{R}_{500})]^2$ and $P_m=(a_m^2+b_m^2)/(2m^2\text{R}_{500}^{2m})$, where $a_0$ is the total intensity within R$_{\text{500}}$ and the moments $a_m$ and $b_m$ are calculated by: $a_m(\text{R})=\int_{\text{R}<\text{R}_{\text{500}}} S(x)\text{R}^m \cos(m \phi)d^2x$ and $b_m(\text{R})= \int_{\text{R}<\text{R}_{\text{500}}} S(x)\text{R}^m \sin(m \phi)d^2x$, where $S(x)$ is the SB at the position x=(R, $\phi$).
In this work we are going to consider the ratios $P_2$/$P_0$ and $P_3$/$P_0$ (hereafter $P_{20}$ and $P_{30}$, respectively), which are respectively a measure of the ellipticity of clusters and of the presence of asymmetries or substructures.
\end{itemize}

\section{Morphological analysis}
\subsection{The observed sample}
\begin{figure}
    \centering
    \includegraphics[scale=0.32]{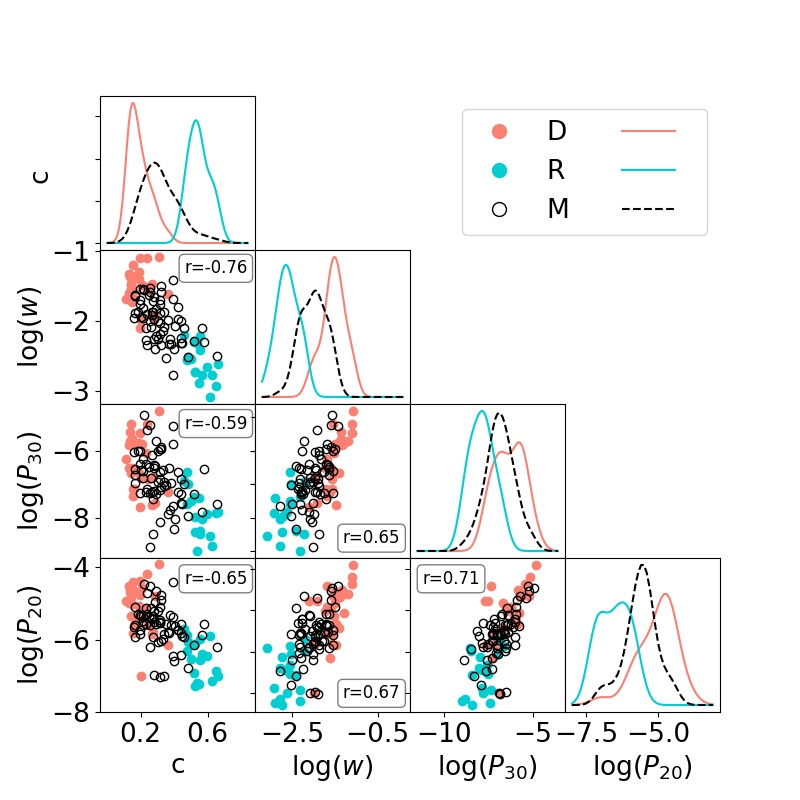}
    \includegraphics[scale=0.31]{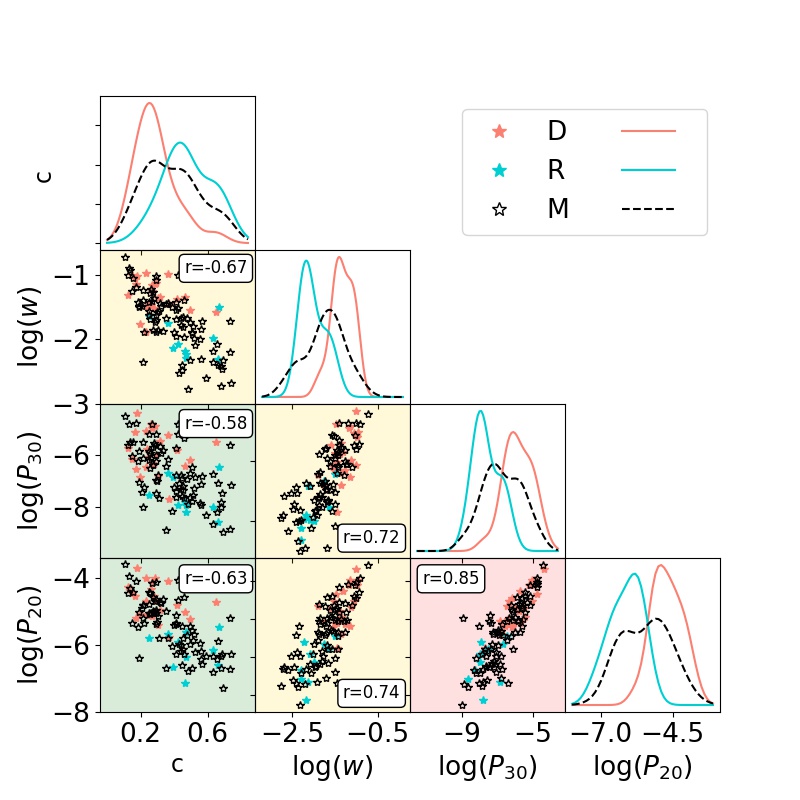}
    \caption{Correlations between the morphological parameters estimated from the observations (left) and simulations (right). Relaxed (R), disturbed (D), and mixed (M) clusters are shown in orange, cyan, and black, respectively. The background colours of the right plot represent the level of confidence between the observed and simulated Spearman coefficient, $r$ (green stays for a level of confidence below 1 $\sigma$, yellow is for a level of confidence between 1 - 3 $\sigma$ and red is for a level of confidence above 3 $\sigma$). }
    \label{fig:corner}
    \vspace{-0.9cm}
\end{figure}

We computed the parameters described above for each cluster and we present the results in Fig. \ref{fig:corner} (left panel) where we also highlight the clusters dynamical state. We  investigated the presence of correlations between the parameters by computing the Spearman coefficient. We found good correlations ($r$ > 0.5) between all parameters with the best (anti-) correlation observed for the couple $c$ -- $w$ ($r$ = -0.76). By using the visual classification of the dynamical state we can see that all parameters are able to separate the relaxed and disturbed populations, with $c$ and $w$ showing the smaller overlap between the distributions.

\subsection{The simulated sample} \label{sec-simulations}
In order to obtain an objective view of the ability of the morphological parameters to separate the relaxed and disturbed populations, we repeat our analysis using a simulated sample provided by {\sc The Three Hundred} collaboration \citep[e.g.,][]{Cui18}. This sample considered in the analysis is composed of 878 clusters spanning a wide range of redshift ($0 < z < 0.6$) and masses (M$_{500}$ > 4.6 $\times$ 10$^{14}$ h$^{-1}$ M$_{\odot}$). The lack of low-mass simulated objects may induce discrepancies between the observed and simulated sample, which will be better investigated in future analysis. In order to reproduce the quality of the \xmm\ observations, we smoothed the simulated images with a Gaussian function of $\sigma$ = 6 arcsec, which represents the FWHM of the \xmm\ PSF, and we binned them using the same scale of the observations (i.e., 1 pixel = 2.5 arcsec).  The great advantage in considering simulations is that it is possible to have a priori knowledge of the dynamical state of clusters.  There are indeed some dynamical indicators that could be used to identify the relaxed and disturbed populations. In our analysis, the indicators considered are the mass fraction of all sub-halo in the cluster, f$_s$, and the offset of the centre of mass, $\Delta_r$, defined respectively as: $F_S$=$\Sigma_i M_i/M_{500}$ and $\Delta_r=|r_{cm}-r_c|/R_{500}$, where M$_{\text{i}}$ is the mass of the sub-halo, r$_{\text{cm}}$ is the centre-of-mass position of the cluster, and r$_{\text{c}}$ is the position of the highest density peak (i.e., the theoretical centre of the cluster). We defined as most relaxed (disturbed) clusters, the systems with both $\Delta_{\text{r}}$ < 0.1 and f$_{\text{s}}$ < 0.1 ($\Delta_{\text{r}}$ > 0.1 and f$_{\text{s}}$ > 0.1, \citep{Cui18, DeLuca2021}). Having determined the dynamical state of the clusters of the simulated sample, we estimated the morphological parameters presented in Section \ref{Sec-morpho}. To compare these new results with the ones obtained for the CHEX-MATE sample, we build 10$^4$ sub-samples of 118 simulated clusters. Each sub-sample is build to reproduce the distribution in redshift and mass of the CHEX-MATE sample. In particular,  we randomly extracted from each redshift snapshot of the simulations a number of objects equal to the number of CHEX-MATE systems located at that redshift. Hereinafter the results shown for the simulated sample are the media of the results obtained for each of the 10$^4$ sub-samples. As first step we compute the level of correlation between the parameter pairs, by determining the Spearman coefficient, and we compare it with the coefficient obtained from simulations. The level of concordance between observations and simulations is reported in Fig. \ref{fig:corner} (right panel): in green are represented the pair of parameters for which the level of agreement is lower than 1$\sigma$, in yellow the pairs for where the level of confidence is included between 1 and 3$\sigma$, and in red the pairs with a level of confidence higher than 3$\sigma$. From this comparison it arises that all the pairs show a level of confidence lower than 3 $\sigma$, with the exception of the P$_{20}$ -- P$_{30}$ whose trend does not reproduce the one arising from the observations.
\section{ROC curves}
To explore the ability of the morphological parameters in detecting the relaxed and disturbed systems we build the so called ROC curves. ROC curves are essentially a comparison between two classifications: the first one, hereafter \textit{real classification}, is based on the dynamical state obtained using the dynamical indicators described in Section \ref{sec-simulations}; the second one, hereafter \textit{morphological classification}, is obtained by defining for each morphological parameter a threshold, V$_T$, above (or below) which, clusters are classified as disturbed (or as relaxed). By comparing the results provided by these two classifications it is possible to define as TP or TN the objects for which the two classifications show agreement, and as FP or FN the objects classified as relaxed from the morphological classification instead as disturbed and vice versa. The ROC curves is then build by varying V$_T$, and computing for each value of V$_T$ the true and the false positive rate (respectively TPR and FPR), defined as: TPR= TP/(TP+FN) and FPR=FP/(FP+TN). By computing the area under the curve (AUC) obtained it is possible to quantify the probability of the considered parameter to correctly identify the dynamical state of clusters: in the perfect case (i.e., 100\% probability of correct detection) the curves obtained is a step function with an AUC = 1. On the contrary, if the probability of the parameter to correctly detect the relaxed and disturbed population is 50\%, the ROC curve obtained is an identity line with AUC = 0.5. By testing this method on our parameters we found that the most powerful parameter is $w$, showing an AUC = 0.89$\pm$0.05, followed by the power ratios $P_{30}$ and $P_{20}$, showing respectively AUC = 0.79$\pm$0.07 and AUC = 0.78$\pm$0.07. For $c$, we found AUC = 0.74$\pm$0.08 suggesting a lower ability to disentangle between the two populations of clusters. 
\section{Study of the systematics}\label{sec-systematics}
\begin{figure}
    \centering
    \includegraphics[scale=0.4]{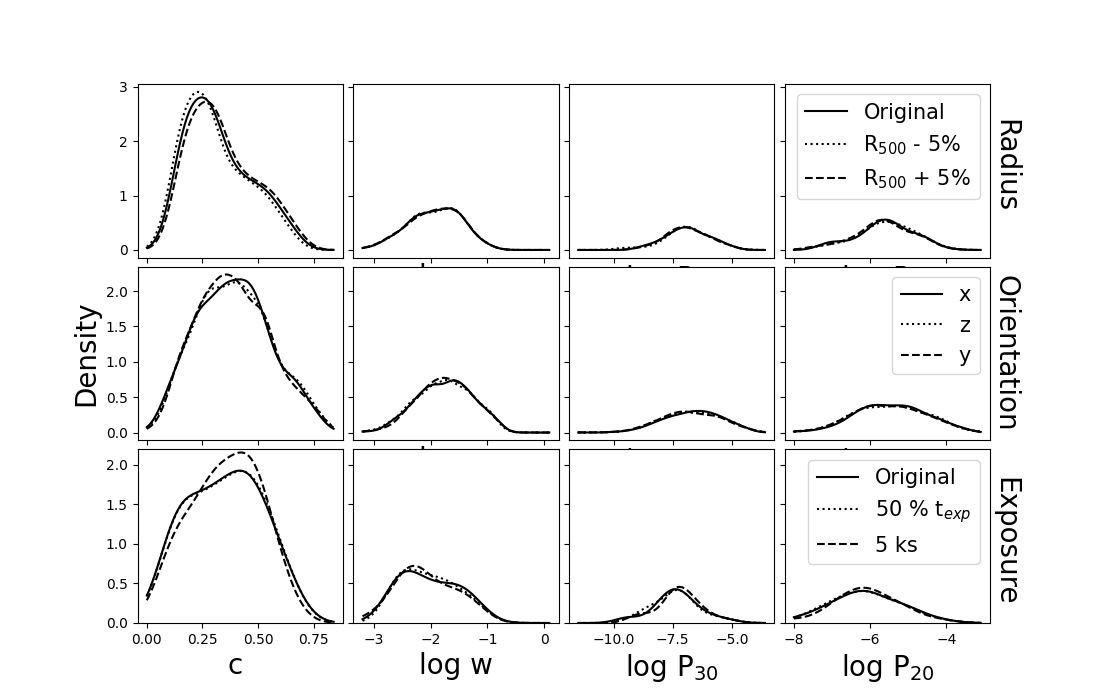}
    \caption{Results of the test reported in Section \ref{sec-systematics}. Top: comparison between the morphological parameters computed in regions with different radii. Center: comparison between the morphological parameters estimated for different orientations of the same cluster. Bottom: comparison between the morphological parameters estimated from original images, images with halved exposure times and images with t$_{\text{exp}}$ $\sim$ 5.}
    \label{fig:tests}
    \vspace{-0.9cm}
\end{figure}
In order to investigate the robustness of the morphological parameters and their dependence on the quality of the images used, we realised the following tests:
\begin{itemize}
    \item decrease of the exposure time: we considered a sub-sample of 20 CHEX-MATE clusters and we repeat the morphological analysis on both images with halved exposure time and images with exposure time equal to 5 ks;
    \item variation the orientation of clusters: we considered the simulated sample which provides for each object three images related to three different orientations (x, y and z). We then compare the values of the morphological parameters estimated along these three different orientations;
    \item change of the radius: we estimate the morphological parameters using two circular regions of of radius r = 1.05 $*$ R$_{500}$ and 0.95 $*$ R$_{500}$. We then compare these two estimations with the original ones (obtained using r = R$_{500}$ ).
\end{itemize}
The results of these tests are reported in Fig. \ref{fig:tests}. The very good agreement between the results indicates that these parameters robustly determine the dynamical state of the clusters.

\section{Combining the morphological parameters}
One of aims of this analysis is to obtain a single quantity which is able to define the grade of relaxation of a system. Therefore, we defined a new parameter $M$ as the combination of the other four parameters \citep{Rasia2013,Cialone2018,DeLuca2021}: $M = \frac{1}{N} \sum A_{Par} \times \frac{Par - m_{Par}}{\sigma_{Par}}$, where N is the number of parameters used, $\sigma$ and $m$ represent respectively the standard deviation and the median of the considered parameter and $A_{Par}$ is equal to -1 for the concentration and 1 for the other parameters. To test the power of this parameter, we first considered the simulated sample. For each of the 10$^4$ sub-samples we followed this procedure: first of all we identify the number of relaxed, f$_r$, and disturbed, f$_d$, clusters as defined by the real classification. Then, we compute $M$ and check whether the first f$_r$ (or f$_d$) objects with the lowest (or highest) values of $M$ are classified as relaxed (or disturbed) by the real classification (so no specific threshold values of $M$ are adopted). If so, we talk about correct detection, otherwise about wrong detection. To estimate the probability of M to correctly or wrongly detect the relaxed objects, we defined the ratio C$_R$ and W$_R$ which are respectively the number of correct or wrong detection of relaxed systems divided for f$_r$ (and similarly C$_D$ and W$_D$ which are the number of correct or wrong detection of disturbed systems divided for f$_d$). We found that the probability to correctly detect relaxed and disturbed systems is C$_R$ = 61 $\pm$ 6 \% and C$_D$ = 54 $\pm$ 8 \%. These percentage are not too high, but could be explained by observing the probability of wrong detection, which is very low, with W$_R$ = 9 $\pm$ 4 and W$_D$ = 10 $\pm$ 5. These latter result suggest that the non-correct detection are mostly due to the mixed population. As it is possible to see in the corner plot of Figure \ref{fig:corner} (right panel) this class of objects is indeed uniformly distributed and could thus alter our results. We then apply the same procedure on the CHEX-MATE sample using as reference the visual classification. We found that the probability of correct detection are C$_R$ = 78 \% and C$_D$ = 73 \%, while the probability of wrong detection are zero for both the relaxed and disturbed populations.
\section{Conclusions}
We realised a morphological analysis of the 118 clusters of the CHEX-MATE sample. In particular, we  investigate the link between the distribution of the X-ray emission and the dynamical state by means of four parameters: $c$, $w$, $P_{20}$, and $P_{30}$. We found that they all show good correlations and are powerful in identifying the most disturbed and most relaxed systems of a sample. Furthermore, we verify that these four indicators are robust and are not strongly influenced by the quality of the images used. We thus combined them in a single quantity, $M$, that allows to build a continuous classification of the grade of relaxation of the systems of the CHEX-MATE sample and we verify that the extreme objects of this classification are effectively relaxed and disturbed systems. 


%
%
%

\end{document}